\documentclass[
twocolumn,
aps,nofootinbib,showpacs,showkeys,preprint
tightenlines,preprintnumbers,
,superscriptaddress
] {revtex4}

\usepackage{epsf,epsfig,subfigure,graphicx,amsmath,amssymb}
\usepackage{color}
\usepackage{float}

\newcommand{\dis}[1]{\begin{equation}\begin{split}#1\end{split}\end{equation}}

 \newcommand{\dell}{\delta_{\rm PMNS}}
\newcommand{\delq}{\delta_{\rm CKM}}

\newcommand\lsim{\mathrel{\rlap{\lower4pt\hbox{\hskip1pt$\sim$}}
    \raise1pt\hbox{$<$}}}
\newcommand\gsim{\mathrel{\rlap{\lower4pt\hbox{\hskip1pt$\sim$}}
    \raise1pt\hbox{$>$}}}

\newcommand\etal{{\it et al.}}

\newcommand\ie{{\it i.e.}~}

\newcommand\gev{\,{\rm GeV}}

\newcommand{\Z}[1]{{\bf Z}}

\begin{document}

\title{Grand unification and CP violation\footnote{Talk presented at ICHEP 2016, Chicago, U.S.A., 3-10 August 2016.}}

\author{Jihn E. Kim }
\affiliation{ 
Department of Physics, Kyung Hee University, 26 Gyungheedaero, Dongdaemun-Gu, Seoul 02447, Republic of Korea, and \\
 Center for Axion and Precision Physics Research (CAPP, IBS),
  291 Daehakro, Yuseong-Gu, Daejeon 34141, Republic of Korea  
}
   
\begin{abstract}
I will talk on my recent works related to the flavor grand unification, weak CP violation, and the phases in the CKM and PMNS matrices.  
\end{abstract}


 \maketitle

\section{Introduction}\label{sect:intro}

\begin{figure}[!b]
  \begin{center}
  \begin{tabular}{c}
   \includegraphics[width=0.45\textwidth]{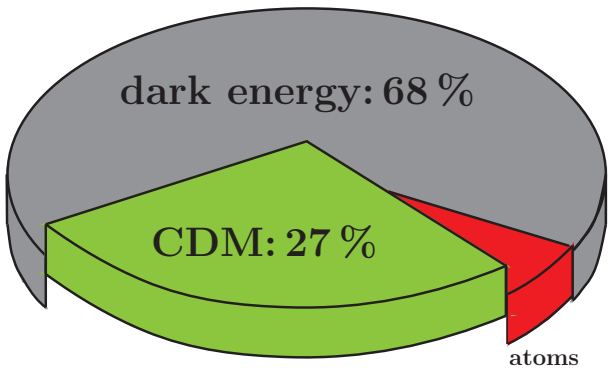}
   \end{tabular}
  \end{center}
 \caption{The energy pie of the Universe.
  }
\label{fig:Epie}
\end{figure}

In this talk, I will concentrate on the mere 5\,\% of the energy pie (mainly of atoms) of the Universe in Fig. \ref{fig:Epie}. In grand unification theories(GUTs), it is the property of chiral fermions, forbidding large masses due to the chiral symmetry. For the flavor solution in GUTs, the GUT symmetry has to be extended. 
Nowadays, flavor symmetries are studied mainly by some discrete symmetries, due to the observed large mixing angles in the leptonic sector. But, flavor symmetry may be a gauge symmetry in which case a true unification is GUTs with the flavor symmetry included there. The first attempt along this line was due to Georgi in SU(11) \cite{Georgi79} on the unification of GUT families (UGUTF). The next try with spinor representation of SO($4n+2$) groups was in SO(14) \cite{KimPRL80}. My attempt along this line in the last year was  from string compactification \cite{KimJHEP15} based on ${\bf Z}_{12-I}$ orbifold compactification \cite{KimKyae07,KimKimKyae07}.
  
\begin{figure}[!t]
  \begin{center}
  \begin{tabular}{c}
   \includegraphics[width=0.45\textwidth]{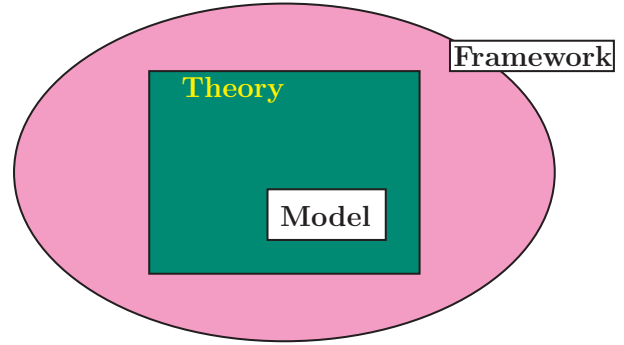}
   \end{tabular}
  \end{center}
 \caption{Gross's thought about the framework-theory-model relation.
  }
\label{fig:Gross}
\end{figure}

Recently, Gross presented a cartoon on our attempts that we are interested in \cite{Gross16}.   There is a grand ``Framework'' which is quite general, acceptable to all scientists. That may include quantum mechanics or symmetry. It may include very interesting ``Theory'' such as Einstein's gravity and string theory, as shown in Fig. \ref{fig:Gross}. Within this theory, one can build a ``Model'' which must be a working example. Even though the design, ``Framework'', is fantastic, without a model example some will say that it is a religion. So, our efforts is to find a working model   toward the theory/framework design. In this vein, we attempt to understand CP violation and GUTs in particle physics and cosmology.
 
To discuss violation of a symmetry, first one has to define the symmetry. Even though kinetic mixings of U(1) gauge bosons have been considered for some time, the definition is usually done such that the kinetic energy terms are diagonal. In the standard model(SM), the kinetic energy terms of quarks, leptons, and Higgs doublets are CP conserving. The CP violation in the SM arises in the interaction terms, typically through the Yukawa couplings. If the VEVs of Higgs doublets vanish, then there is no CP violation because all fermions are massless. Below the VEV scale of the Higgs doublets, all the SM fields obtain masses, and one can locate the CP phase in the left-handed currents, coupling to $W^\pm_\mu$.  The charged current couplings are defined in this setup and  the CKM and PMNS matrices are defined respectively for quarks and leptons.
The CKM and PMNS matrices are unitary, which is the only condition for the CKM and PMNS matrices, and so there are many different parametrization schemes \cite{PMNS,KM73,CK84,Maiani76,KimSeo11}. Because the CKM matrix elements are rather well-known by now, there are three classes of parametrizations \cite{KimNam15}, each having the same CP phase $\delq$.

The CP violation in the SM is an interference phenomenon, encompassing all three families. This will become clearer below when we express the Jarlskog determinant $J$.

\section{Jarlskog phases in the CKM and PMNS matrices}

The discussion on the strong CP is not separable from the discussion of the weak CP violation. Nowadays, the strong CP problem is well understood in ``invisible'' axions \cite{KimRMP10}. So, I will concentrate on the weak CP here.

  Recently, it has been pointed out that a new parametrization of the CKM matrix $V_{\rm CKM}~(\equiv V~\rm below)$ with one row (or column) real is very useful to scrutinize the physical effects of the weak CP violation. Then, the elements of the determinant directly give  the weak CP phase \cite{KimSeo11}. The physical significance of the weak CP violation is given by the  Jarlskog determinant $J$ \cite{Jarlskog85} which is obtained from the imaginary part of a product of two elements of $V$ and two elements of $V^*$ of the CKM matrix, e.g. of the type $V_{12}V_{23}V_{13}^*V_{22}^*$. This Jarlskog determinant is just twice the area of the Jarlskog triangle. In Ref.  \cite{KimSeo12}, we have shown that one easily obtains the Jarlskog determinant from the entries of $V$ if Det.$V=1$.  If Det.$V\ne 1$, one can make it so by multiplying an overall phase to all the up- or all the down-type quarks. For example, the  Jarlskog determinant $J$ is the imaginary part of the product of the skew diagonal elements, $J=|{\rm Im\,}V_{13}V_{22}V_{31}|$. To relate the product of four elements of $V$ and $V^*$ to a product of three elements of $V$ can be proved as follows. If the determinant of $V$ is real, we have ~$ 1=
 V_{11}V_{22}V_{33}-V_{11}V_{23}V_{32} +V_{12}V_{23}V_{31}
 -V_{12}V_{21}V_{33} +V_{13}V_{21}V_{32} -V_{13}V_{22}V_{31}.
$
Multiplying $V_{13}^*V_{22}^*V_{31}^*$ on both sides, we obain
\dis{
V_{13}^* &V_{22}^*V_{31}^*  =|V_{22}|^2V_{11}V_{33}V_{13}^*V_{31}^*-V_{11}
 V_{23}V_{32}V_{13}^*V_{31}^*V_{22}^*\\
 & +|V_{31}|^2V_{12}V_{23}V_{13}^*V_{22}^* -V_{12}V_{21}V_{33}V_{13}^*V_{31}^*
 V_{22}^* \\
 &+|V_{13}|^2V_{21}V_{32}V_{31}^*V_{22}^*-|V_{13}V_{22}V_{31}|^2.\label{eq:detmultV*}
}
 Let the imaginary part of $V_{11}V_{33}V_{13}^*V_{31}^*$ be $J$. Then, Im\,$V_{13}^*  V_{22}^*V_{31}^*$ is  Im\,$V_{11}V_{33}V_{13}^*V_{31}^*$, \ie  $J$= Im\,$V_{13}^*  V_{22}^*V_{31}^*$  \cite{KimMoNam15}.

\begin{figure}[!b]
  \begin{center}
  \begin{tabular}{c}
   \includegraphics[width=0.3\textwidth]{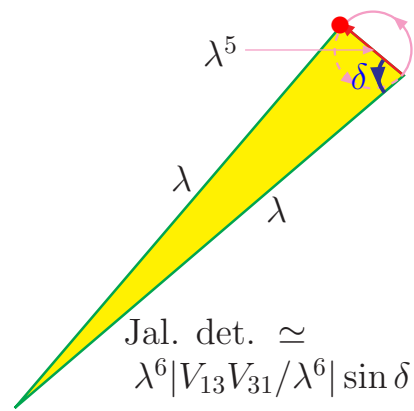}
   \end{tabular}
  \end{center}
 \caption{The Jarlskog triangle. This triangle is for two long sides of $O(\lambda)$. Rotating the $O(\lambda^5)$ side (the red arrow), the CP phase $\delq$ changes.
  }
\label{fig:JTriangle}
\end{figure}

By looking at the above triple product form on $J$, we can see where the  physical CP phase appears, and in terms of the Kim-Seo(KS) parametrization angles \cite{KimSeo11},
\dis{
J=|c_1c_2c_3s_1^2s_2s_3  \sin\delq|,
}
where $s_1=\sin\theta_1$, etc. The above form of $J$ includes all mixing angles, \ie all three families must parcitipate in the diagram for weak CP violation. CP violation is an interference phenomenon.
  Note that the unitarity triangle in the PDG book gives \cite{PDG15}
\dis{
\textrm{PDG15:}~ &\alpha=\left(85.4^{+3.9}_{-3.8}\right)^{\rm o},
~\beta=\left(21.50^{+0.76}_{-0.74}\right)^{\rm o}\\
&\gamma=\left( 68.0^{+8.0}_{-8.5}\right)^{\rm o}.\label{eq:PDGfit}
 }
 The recent UTfit gives  \cite{Pich16},
 \dis{
\textrm{UTfit16:}~&\alpha=\left(88.6\pm3.3
\right)^{\rm o},~\beta=\left( 22.03\pm 0.86\right)^{\rm o},\\
&\gamma=\left( 69.2\pm 3.4\right)^{\rm o},
\label{eq:UTfit}
 }
and the recent CKMfit gives with the unitarity constraint over the world average \cite{HarnewH16} 
  \dis{
\textrm{CKMfit16:}~&\alpha=\left(90.6^{+3.
9}_{-1.1}
\right)^{\rm o},~\beta=\left( 24.21^{-1.33}_{-1.35}\right)^{\rm o},\\
&\gamma=\left( 66.9^{+0.94}_{-3.44}\right)^{\rm o}.\label{eq:CKMfit}
 }

So, there are three possibilities for the CP phase from our formula on $J$: $\delq=\alpha,\beta,$ or $\gamma$. Note that $\alpha$ is close to $\frac{\pi}{2}$ in Eqs. (\ref{eq:PDGfit},\ref{eq:UTfit},\ref{eq:CKMfit}). Make Det=1 as in the KS form \cite{KimSeo11}.  Then, we observe that
\begin{itemize}
\item[1.]
Make the real part of (22) element  very large as in many parametrizations.        Then,
\dis{
\delq=\alpha.\label{eq:phasealpha}
}
The parametrizations in Ref. \cite{KimSeo11} give $\delq=\alpha$. Also, the Kobayashi-Maskawa form \cite{KM73} gives (\ref{eq:phasealpha}), by multiplying a universal phase since its determinant is not real.

\item[2.] 
      If both 1st row = real and 1st column = real, then the Chau-Keung parametrization \cite{CK84} gives
\dis{
\delq=\gamma.\label{eq:phasegamma}
}
Also, the Maiani  parametrization \cite{Maiani76} gives (\ref{eq:phasegamma}), by multiplying a universal phase.
\end{itemize}
 So, it is proved that $J=                                  
|{\rm Im.}\,V(31) V(22) V(13) |$ is very useful.                                    

We can argue that the maximality of the weak CP violation is a physical statement. The physical magnitude of the weak CP violation is given by the area of the Jarlskog triangle. For any Jarlskog triangle, the area is the same. With the $\lambda=\sin\theta_C$ expansion, the area of the Jarlskog triangle is of order $\lambda^6$. In Fig. \ref{fig:JTriangle}, we show the triangle with two long sides of order $\lambda$. Rotating the $O(\lambda^5)$ side (the red arrow of Fig. \ref{fig:JTriangle}), the CP phase $\delta$ and also the area change. The magnitude of the Jalskog determinant is $J\simeq\lambda^6|V_{13}V_{31}/\lambda^6|\sin\delta$. From  Fig. \ref{fig:JTriangle}, we note that the area is maximum for  $\delta\simeq\frac{\pi}{2}$, and the maximality $\delta=\frac\pi{2}$ is a physical statement. In the above, the maximal CP violation was proved very easily with the KS parametrization \cite{KimSeo11}. The same must be true even if we use the CKM parametrization \cite{CK84,Maiani76}, but the proof on the maximality around the observed real parameters may not be so simple.

Within this scheme, we cannot determine which parametrization is more useful over the others. There must be another independent condition to choose a proper parametrization.

\section{One phase in 
the theory?}

\begin{figure}[!t]
  \begin{center}
  \begin{tabular}{c}
   \includegraphics[width=0.45\textwidth]{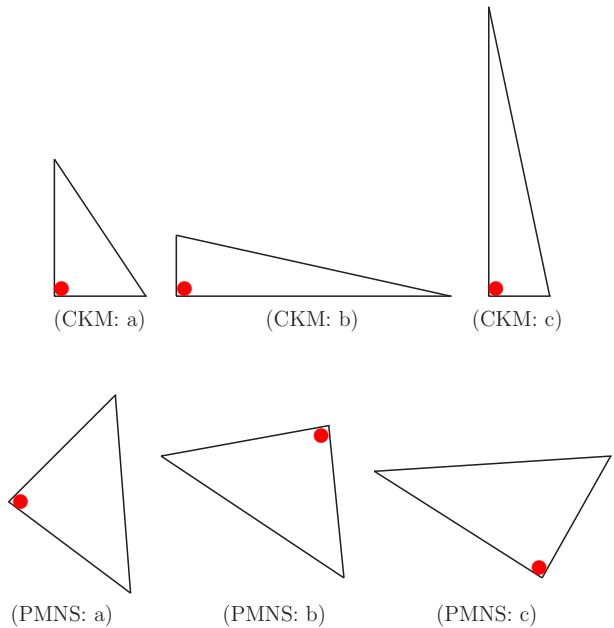}
   \end{tabular}
  \end{center}
 \caption{A schematic view of the Jarlskog triangles of quarks and leptons. If the quark and lepton mixing phases are related, the Jarlskog triangles may have these shapes, with the same angle appearing in all the triangles.
  }
\label{fig:qandlphase}
\end{figure}

In the previous section, we observed that there is a possibility that $\delq=\frac{\pi}{2}$. In the leptonic sector also, there is a preliminary hint that $\dell\ne 0$, and close to $-\frac{\pi}{2}$ even though the error bar is large \cite{T2KCP}. The quark and lepton mixing angles can be parametrized by
 \dis{
&\left(
\begin{array}{ccc}
c_1 & s_1c_3 & s_1s_3  \\
-c_2s_1 & -e^{-i\delta'}s_2s_3+c_1c_2c_3 & e^{-i\delta'}s_2c_3+c_1c_2s_3  \\
e^{i\delta'}s_1s_2 & -c_2s_3-c_1s_2c_3 e^{i\delta'} & c_2c_3-c_1s_2s_3e^{i\delta'} \\
\end{array}\right),\\[0.2em]
 &\left(
\begin{array}{ccc}
C_1 & S_1C_3 & S_1S_3  \\
-C_2S_1 & -e^{-i\delta}S_2S_3+C_1C_2C_3 & e^{-i\delta}S_2C_3+C_1C_2S_3  \\
e^{i\delta}S_1S_2 & -C_2S_3-C_1S_2C_3 e^{i\delta} & C_2C_3-C_1S_2S_3e^{i\delta} \\
\end{array}\right)\nonumber
 } 
where $\delta'=\delq$ and $\delta=\dell$ \cite{PMNS}, and the lower and upper cases of sine and cosine of angles $\theta_i$ and $\Theta_i$ are for $q$ and $l$, respectively, \ie $s_i=\sin\theta_i$ and $S_i=\sin\Theta_i$. Even if  $\theta_i$ and $\Theta_i$ cannot be related, we can relate $\delq$ and $\dell$ if there is only one CP phase in the whole theory. Indeed, this has been shown in Ref. \cite{KimPLB11} where the weak CP violation is spontaneous and one unremovable phase is located at the weak interaction singlet {\it \`a la} the Froggatt-Nielsen(FN) mechanism \cite{FN79}. In the supersymmetric model, it was shown that one phase in the ultra-violet completion gives \cite{KimNam15}
\dis{
\dell=\pm\delq.
}
Then, the Jarlskog triangles of the quark and lepton sectors will look like Fig. \ref{fig:qandlphase} where the same phase appears in all the diagrams. Of course, the real angles in the quark and lepton sectors are different.
  
There are many phases in concern: CKM, PMNS, Majorana, and leptogenesis phases. 
If there is only one phase, all of these must be expressed in terms of one phase.  So, the Majorana phase determined at the intermediate scale and the leptogenesis phase can be also expressed in terms of this one phase, as shown in Ref. \cite{CoviKim16}.

One obvious strategy to relate the CKM and PMNS phases is GUTs, but the CP phase is a property of families. This leads us to the consideration of UGUTF as commented in Introduction. There are the constraint from the
FCNC, \ie the family symmetry breaking scale is above $10^5\,\gev$.  

\begin{figure}[!t]
  \begin{center}
  \begin{tabular}{c}
   \includegraphics[width=0.45\textwidth]{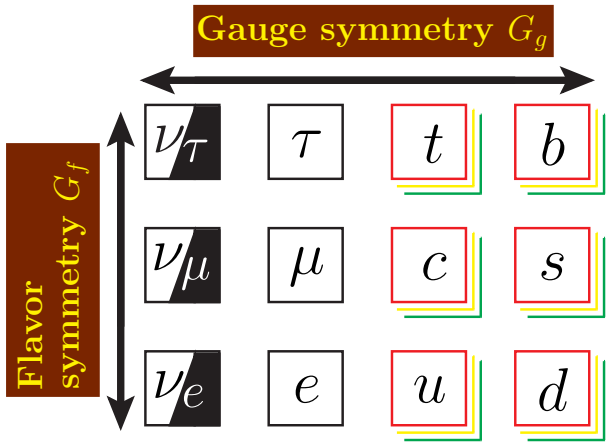}
   \end{tabular}
  \end{center}
 \caption{A cartoon for $G_g\times G_f$ with three colors of quarks.}\label{fig:GgGf} 
\end{figure}

In a talk in the flavor physics parallel session here, theories on mixing angles were classified as \cite{Zhou16}
\begin{itemize}
\item[(i)]   SM $\times$ (family symmetry), 
\item[(ii)]  GUT $\times$ (family symmetry),  
\item[(iii)] Unification of GUT families in
      a simple gauge group,
\end{itemize}
where the family symmetry was considered as a discrete group such as $A_4$.
 In Fig.  \ref{fig:GgGf}, the direct product of gauge and flavor symmetries are shown. In Table \ref{tab:DiscCont},  the flavor group $G_f$ is further classified depending on discrete and continuous possibilities. These classifications belong to Items (i) and (ii). In the remainder of my talk, however, I concentrate on Item (iii), where $G_g\times G_f$ is in a simple group.
\begin{table}[h]
\begin{center}
\begin{tabular}{| c|c|c| }
\hline 
    &Continuous&Discrete  \\ 
  \hline  
Abelian&U(1)&  {\bf Z}$_N$ \\
\hline
Non-Abelian& SU(3), SO(3),$\cdots$ &  $A_4,S_4,A_5,\Delta(48),\cdots$   \\ 
 \hline 
  \end{tabular}
\end{center}
\caption{A classification of $  G_f$ \cite{Zhou16}.}\label{tab:DiscCont}
\end{table}

As commented in Introduction, we discuss 
UGUTFs. In string compactification, there is one example SU(7)$\times$U(1) from {\bf Z}$_{12-I}$ orbifold compactification \cite{KimJHEP15}, and the possibility for one CP phase is also obtained in \cite{KimNam15}. 

\subsection{UGUTFs}
In the unification of families in SU($N$) GUTs, SU(5) families are counted by the number of {\bf 10}\,s \cite{Georgi79}. For example, a few antisymmetric representations of SU($N$) allow the following family numbers,
\dis{
&\textrm{SU(5)}:\,\Psi^{[\alpha\beta]},\,n_f=1;\\
&\textrm{SU(6)}:\,\Psi^{[\alpha\beta\gamma]},\,n_f=0,~~
\Psi^{[\alpha\beta]},\,n_f=1; \\
&\textrm{SU(7)}:\,\Psi^{[\alpha\beta\gamma]},\,n_f=1,~~
\Psi^{[\alpha\beta]},\,n_f=1.
}
With this kind of counting, Ref. \cite{Georgi79} in SU(11) with $\Psi^{[\alpha\beta\gamma\delta]}\oplus \Psi_{[\alpha\beta\gamma]}\oplus \Psi_{[\alpha\beta]} \oplus\Psi_{[\alpha]}$ obtained three SU(5) families. There can be numerous examples in SU($N$) if we allow repetition of some representations \cite{Frampton79}. In string compactification, it is very difficult to obtain  large SU($N$) groups. For example, the rank 8 group E$_8$ cannot allow a subgroup with rank greater than 8. Reference \cite{KimJHEP15} obtained a UGUTFs of three families in SU(7)$\times$U(1), a kind of flipped form. The SU(7) representations are a $\Psi^{[\alpha\beta\gamma]}$ from the untwisted sector and two $\Psi^{[\alpha\beta]}$\,s from the twisted sector. Thus, it gives three families.

\subsection{One phase}
In the model of  \cite{KimJHEP15}, it has been shown that there is a possibility of $\dell=\pm\delq$ \cite{KimMoSeo15} if only one CP phase is present in the GUT scale VEVs. Since the discussion is more involved there, we explain in a simpler study obtained from the model of \cite{KimPLB11}. The up-type quark and the down-type quark mass matrices are represented as
\dis{
&\quad\quad\quad\quad\quad u_R(+5) \quad   c_R(+4) \quad  t_R(+2) \\[-0.2cm]
&\quad\quad\quad\quad-------------\\[-0.2cm]
\tilde{M}^{(u)}=&\begin{pmatrix} 
\bar{q}(+1)&|&cX_{-1}^{u\,6}& -cX_{-1}^{u\,5}&\kappa_tX_{-1}^{u\,3}\\
\bar{q}(0)&|&-cX_{-1}^{u\,5}& -cX_{-1}^{u\,4}&-\kappa_tX_{-1}^{u\,2}\\
\bar{q}(-2)&|&\kappa_t X_{-1}^{u\,3}&  -\kappa_tX_{-1}^{u\,2}&1 
\end{pmatrix}  v_u,\label{eq:u}
}
for the dimensionless up-type quark mass matrix, and
\dis{
&\quad\quad\quad\quad\quad d_R(-5) \quad ~  s_R(0) \quad \quad~  b_R(+2) \\[-0.2cm]
&\quad\quad\quad\quad-------------\\[-0.2cm]
\tilde{M}^{(d)}=&\begin{pmatrix} 
\bar{q}(+1)&|&dX_{+1}^{d\,4}& -0&0\\
\bar{q}(0)&|&0& -sX_{+1}^{d}X_{-1}^{d}&-\kappa_bX_{-1}^{d\,2}\\
\bar{q}(-2)&|&0&  -\kappa_bX_{+1}^{d\,2}&1 
\end{pmatrix}  v_d\label{eq:d}
}
for the dimensionless down-type quark mass matrix. Here, $q$'s are the left-handed fields, $u_R,d_R,\cdots$, are the right-handed fields, and their {\bf Z}$_{12}$ quantum numbers are shown inside brackets. The fields $X$'s are the FN fields with  {\bf Z}$_{12}$ quantum numbers shown as subscripts. For the spontaneous CP violation, entries must have different phases, realized by different powers of $X^{(u),(d)}$, such that there is no way to remove the phases completely. Except the phases of $X$ fields, all parameters are real.  Under certain assumptions on the superpotential on $X^{(u)}$ and $X^{(d)}$, we obtain the phase $e^{i\,\frac{\pi}2}$ for  $X^{(d)}$. In this FN type form, we showed Eqs. (\ref{eq:u}) and (\ref{eq:d}) just to show how the entries must have different phases toward an unremovable phase in the end.
 
 \vskip 0.5cm
\section{Conclusion}

My talk on weak CP is centered on flavor unification, emphasizing the possibility in GUTs. A few emphases were:
\begin{itemize}
\item[(1)] The Jarlskog determinant $J$ is $|{\rm Im\,} V_{31} V_{22} V_{13}|$ in the KS form,  
\item[(2)] It is shown that the Jarlskog determinant $J$ is almost maximum with the current determination of quark (real) mixing angles. There is a possibility that $\delq$ can be $\frac{\pi}{2}$, which is the case in the KS \cite{KimSeo11} and Kobayashi-Maskawa \cite{KM73} forms of the CKM matrix, 
\item[(3)]  There is a possibility that $\dell$ can be maximal \cite{T2KCP}, 
\item[(4)]   Unification of  GUT families is possible with SU(7)$\times$U(1), which is derived in {\bf Z}$_{12-I}$ orbifold compactification \cite{KimJHEP15}, and 
\item[(5)]  With the FN singlets it is possible to relate the  CKM, PMNS, Majorana, and leptogenesis phases.
\end{itemize}

\acknowledgments{This work is supported in part by the National Research Foundation (NRF) grant funded by the Korean Government (MEST) (NRF-2015R1D1A1A01058449) and  the IBS (IBS-R017-D1-2016-a00).}


\def\prp#1#2#3{{Phys.\,Rep.}  {\bf #1} (#3) #2}
\def\rmp#1#2#3{{Rev. Mod. Phys.}  {\bf #1} (#3) #2}
\def\npb#1#2#3{{ Nucl.\,Phys.\,B}   {\bf #1} (#3) #2}
\def\plb#1#2#3{{Phys.\,Lett.\,B}   {\bf #1} (#3) #2}
\def\prd#1#2#3{{Phys.\,Rev.\,D}   {\bf #1} (#3) #2}
\def\prl#1#2#3{{Phys.\,Rev.\,Lett.}   {\bf #1} (#3) #2}
\def\jhep#1#2#3{{JHEP}   {\bf #1} (#3) #2}
\def\jcap#1#2#3{{JCAP}   {\bf #1} (#3) #2}
\def\zp#1#2#3{{Z.\,Phys.}   {\bf #1} (#3) #2}
\def\njp#1#2#3{{New\,J.\,Phys.}   {\bf #1} (#3) #2}
\def\frp#1#2#3{{Front.\,Phys.}    {\bf #1}  (#3) #2}
\def\epjc#1#2#3{{Euro.\,Phys.\,J.\,C}    {\bf #1} (#3) #2}
\def\anp#1#2#3{{Ann.\,Phys.}    {\bf #1} (#3) #2}
\def\jpg#1#2#3{{J.\,Phys.\,G}   {\bf #1} (#3) #2}
\def\ijmpd#1#2#3{{Int.\,J.\,Mod.\,Phys.\,D}   {\bf #1} (#3) #2}
\def\mpla#1#2#3{{Mod.\,Phys.\,Lett.\,A}   {\bf #1} (#3) #2}
\def\apj#1#2#3{{Astrophys.\,J.}    {\bf #1} (#3) #2}
\def\nat#1#2#3{{Nature}    {\bf #1} (#3) #2}
\def\sjnp#1#2#3{{Sov.\,J.\,Nucl.\,Phys.}   {\bf #1} (#3) #2}
\def\apj#1#2#3{{Astrophys.\,J.}   {\bf #1} (#3) #2}
\def\mnra#1#2#3{{Mon.\,Not.\,Roy.\,Astron.\,Soc.}    {\bf #1} (#3) #2}
\def\jetpl#1#2#3{{JETP\,Lett.}   {\bf #1} (#3) #2}
\def\prth#1#2#3{{Prog.\,Theor.\,Phys.}    {\bf #1} (#3) #2}
\def\jkps#1#2#3{{J.\,Korean\,Phys.\,Soc.}   {\bf #1} (#3) #2}
\def\dum#1#2#3{{\bf #1}, #2 (#3)}

\def\ibid#1#2#3{{\it ibid.}   {\bf #1} (#3) #2}
\def\err#1#2#3{ {\bf #1}  {\bf #1} (#3) #2\,(E) (#3)}   


\end{document}